\documentstyle[11pt,epsf,graphicx]{article}
\def\diff{{\rm\,d}}                    %simbolo di derivata totale
\def\r{\mbox{\boldmath $r$}}

\def\q{\mbox{\boldmath $q$}}
\def\mcg{\mbox{$\mathcal{G}$}}
\def\mcv{\mbox{$\mathcal{V}$}}
\begin{document}

\begin{center}
{\Large\bf{Green's Function Approach to Inclusive Electron Scattering}
\footnote[1]{presented by C. Giusti, E-mail: giusti@pv.infn.it, 
phone: +39 0382507454, fax: +39 0382526938.}}
\end{center}
\vskip1.0truecm
\centerline{\large{F. Capuzzi, C. Giusti, A. Meucci and F.D.
Pacati}}
\vskip1.0truecm
\centerline{\small{Dipartimento di Fisica Nucleare e Teorica, 
Universit\`a degli Studi di Pavia, and}}

\centerline{\small{Istituto Nazionale di Fisica Nucleare, Sezione di
Pavia, Pavia, Italy}}

\begin{abstract}
A Green's function approach to the inclusive quasielastic ($e,e'$) 
scattering is presented. The components of the nuclear response are written in
terms of the single-particle optical model Green's function. The explicit
calculation of the Green's function can be avoided by its spectral
representation, which is based on a biorthogonal expansion in terms of the 
eigenfunctions of the non-Hermitian optical potential and of its Hermitian
conjugate. This allows one to treat final state interactions consistently in 
the inclusive ($e,e'$) and in the exclusive ($e,e'N$) reactions. Numerical 
results for the longitudinal and transverse response functions obtained in a 
nonrelativistic and in a relativistic framework are presented and discussed 
also in comparison with data.  
\end {abstract}
\vspace{1cm}

\section{Introduction}
Inelastic electron scattering has proven to be a very precise tool for studying
the excitation spectrum of atomic nuclei owing to the possibility of
simultaneously varying energy and momentum transfer ($\omega,\q$) 
\cite{Oxford}. In a representation of the nuclear response as a function of 
$\omega$ and $q$ a large broad peak occurs at about $\omega=q^2/2M$. Its 
position corresponds to the elastic peak in electron scattering by a free 
nucleon. It is quite natural to assume that a quasifree process is reponsible 
for such a peak with a nucleon emitted quasielastically. Coincidence ($e,e'N$) 
experiments in the quasifree region confirm such a picture and represent a 
valuable source of information on single-nucleon degrees of freedom inside 
nuclei. 

In the inclusive ($e,e'$) process only the scattered electron is detected and 
the final nuclear state is undetermined, but the main contribution in the 
region of the quasielastic (QE) peak comes from the interaction on single
nucleons. If the nucleons were indeed free, the peak would be sharp and would
just occur at the energy loss  $\omega=q^2/2M$ corresponding to the energy taken
by the recoiling free nucleon. A shift in the position of the peak is
produced by the nuclear binding, while a broadening of the peak is produced by
Fermi motion.  A very simple model, where the cross section is given by the
interaction on all the constituent nucleons of the nucleus and a Fermi gas model
is assumed for nuclear structure, is indeed able to give a remarkably good
description of the cross section in the QE region \cite{Moniz}. 
This same model, however, is not able to describe simultaneously the QE
longitudinal and transverse responses, whose Rosenbluth separation
has been experimentally achieved on a variety of nuclei and over a large range
of momentum transfers \cite{Oxford}.    

In the one-photon exchange approximation and assuming the plane-wave
approximation for the incident and the outgoing electrons, the ($e,e'$) cross 
section is given by \cite{Oxford}
\begin{equation}
\sigma_{\rm{inc}} = K \left( 2\varepsilon_{\rm{L}}R_{\rm{L}}
 + R_{\rm{T}}\right) \ , \label{eq.cs}
\end{equation} 
where $K$ is a kinematical factor and 
\begin{equation}
\varepsilon_{\rm{L}} = \frac{Q^2}{\q^2} \left( 1 + 2 \frac{\q^2}{Q^2}
\tan^2{(\vartheta_e/2)}\right)^{-1} \ \label{eq.polar}
\end{equation}
measures the polarization of the virtual photon. In Eq. (\ref{eq.polar})
$\vartheta_e$ is the scattering angle of the electron, $Q^2 = \q^2 - \omega^2$,
and $q^{\mu} = (\omega,\q)$ is the four momentum transfer. 
The longitudinal and transverse response functions, $R_{\rm{L}}$ and 
$R_{\rm{T}}$ are defined by
\begin{equation}
R_{\rm{L}}(\omega,q) = W_{\rm{tot}}^{00}(\omega,q),  
R_{\rm{T}}(\omega,q) = W_{\rm{tot}}^{xx}(\omega,q) + 
      W_{\rm{tot}}^{yy}(\omega,q) \ ,
\label{eq.response}
\end{equation}
in terms of the diagonal components of the hadron tensor
\begin{equation}
W^{\mu\mu}_{\rm{tot}}(\omega,q) = 
 \int \sum_{\rm f} \mid \langle 
\Psi_{\rm f}\mid J^{\mu}(\q) \mid \Psi_0\rangle\mid^2 
\delta (E_0 +\omega - E_{\rm f}) \ . \label{eq.hadrontensor}
\end{equation}
Here $J^{\mu}$ is the nuclear charge-current operator which connects the 
initial state $\mid\Psi_0\rangle$ of the nucleus, of energy $E_0$, with the 
final states $\mid \Psi_{\rm {f}} \rangle$, of energy $E_{\rm {f}}$. 
In the inclusive ($e,e'$) process the sum is over a complete set of final
nuclear states and only the diagonal components of the hadron tensor contribute,
while in the exclusive ($e,e'N$) process also the nondiagonal components
contribute and the cross section is in general given in terms of a larger 
number of structure functions \cite{Oxford}. 

The separation of the response functions $R_{\rm{L}}$ and $R_{\rm{T}}$ 
can experimentally by achieved from Eq. (\ref{eq.cs}) by varying electron
kinematics. In a plot of the cross section as function 
$\varepsilon_{\rm{L}}$,  for fixed values of $\omega$ and $Q^2$, the slope
gives $R_{\rm{L}}$ and the extrapolated intercept with the vertical axis at 
$\varepsilon_{\rm{L}}=0$ gives $R_{\rm{T}}$. Such a Rosenbluth 
separation \cite{Rosen} has been achieved in the 80's and 90's on different 
nuclei \cite{Oxford}. 

A huge amount of theoretical work was produced over the past two decades in 
order to explain the problem raised by the experimental separation of 
$R_{\rm{L}}$ and $R_{\rm{T}}$ \cite{Oxford}. 
The main problem was the apparent quenching of $R_{\rm{L}}$, while for 
$R_{\rm{T}}$ there is in general an apparent excess of strength.  Different
approaches, with different theoretical ingredients, obtained partial success in
explaining either the longitudinal or the transverse response, but despite all 
these efforts some problems are still open and a consistent and simultaneous
description of $R_{\rm{L}}$ and $R_{\rm{T}}$ data has not been achieved.
Some confusion is also due to the experimental situation. These experiments of
separation are difficult and data from different laboratories are sometimes in
disagreement. New experiments with improved experimental accuracy would be
helpful to make the experimental situation clearer. New measurements are planned
at Jlab \cite{jlabpro}

In this contribution we present a Green's function approach to the inclusive 
QE ($e,e'$) scattering. With our work we don't aim at a unified and 
consistent description of $R_{\rm{L}}$ and $R_{\rm{T}}$ data. The model
contains approximations. It is basically a single-particle approach, where 
only the one-body part of the nuclear current is retained. Calculations of 
two-body contributions performed by different groups have given different 
results \cite{Oxford,Sick}. There are, however, consistent 
indications that the combined effect of two-body current and tensor 
correlations can be appreciable on the transverse response 
\cite{Leidemann, Fabrocini,Sick}. Thus, with our model we don't expect to 
describe $R_{\rm{T}}$ data, while the main contributions to $R_{\rm{L}}$ should 
already be included. Our aim is to study the effects of final state 
interactions (FSI). FSI are an important ingredient of the inclusive electron 
scattering, since they are essential to explain the exclusive one-nucleon 
emission, which gives the main contribution to the inclusive reaction in the QE 
region. The absorption in a given final state due, e.g., to the imaginary part 
of the optical potential, produces a loss of flux that is appropriate for the 
exclusive process, but inconsistent for the inclusive one, where all the 
allowed final channels contribute and the total flux must be conserved. 

This conservation is preserved in the Green's function approach, where the 
components of the nuclear response are written in terms of the single-particle 
optical model Green's function. This result was originally derived by
arguments based on the multiple scattering theory \cite{hori} and successively
by means of the Feshbach projection operator formalism 
\cite{chinn,bouch,capuzzi,capma}. The spectral representation of the
single-particle Green's function, based on a biorthogonal expansion in terms 
of the eigenfunctions of the non-Hermitian optical potential, allows one to 
perform explicit calculations and to treat FSI consistently in the inclusive 
and in the exclusive reactions. Important and peculiar effects are given, in 
the inclusive reactions, by the imaginary part of the optical potential, which 
is responsible for the  redistribution of the strength among different 
channels. The approach has been developed and used in a nonrelativistic frame
\cite{capuzzi} and, more recently, also in a relativistic frame \cite{rel}. 
Although some differences and complications are due to the Dirac matrix 
structure, the formalism follows in both frames the same steps and 
approximations. The numerical results obtained in both frames 
allow us to check the relevance of relativistic effects.

The Green's function approach is presented in  Sec. \ref{sec.green}.  
In Sec. \ref{sec.results} some results for $R_{\rm{L}}$ and 
$R_{\rm{T}}$ are presented and discussed also in comparison with data. 
Summary and conclusions are drawn in Sec. \ref{sec.conc}.

\section{The Green's function approach}
\label{sec.green}

The response functions of the inclusive ($e,e'$) scattering are defined in 
Eq. (\ref{eq.response}) in terms of the diagonal components of the
hadron tensor of Eq. (\ref{eq.hadrontensor}), that can equivalently be 
expressed as
\begin{equation}
W^{\mu\mu}_{\rm{tot}}(\omega,q) = -\frac{1}{\pi} {\rm{Im}} \langle \Psi_0
\mid J^{\mu\dagger}(\q) G(E_{\rm {f}}) J^{\mu}(\q) \mid \Psi_0 \rangle \
,\label{eq.hadrten}
\end{equation}
where $E_{\rm {f}} = E_0 +\omega$ and $G(E_{\rm {f}})$ is the Green's
function related to the nuclear Hamiltonian $H$, i.e.,
\begin{equation}
G(E_{\rm {f}}) = \frac{1}{E_{\rm {f}} - H + i\eta} \ . \label{eq.green}
\end{equation}
Here and in all the equations involving $G$ the limit for $\eta \rightarrow
+0$ is understood. It must be performed after calculating the matrix elements 
between normalizable states.

The hadronic tensor in Eq. (\ref{eq.hadrten}) contains the full (A+1)-body
propagator of the nuclear system and, as such, it is an extremely complicated
object, which defies a practical evaluation. Only an approximated treatment 
reduces the problem to a tractable form. Under suitable approximations the
nuclear response can be written in terms of the optical-model Green's function 
\cite{capuzzi}. 

The first approximation consists in retaining only the one-body part of the 
charge-current operator $J^{\mu}$. Thus, we set
\begin{equation}
J^{\mu}(\q) = \sum_{i=1}^{A+1} j_i^{\mu}(\q) \ , \label{eq.1bcurrent}
\end{equation}
where $j_i^{\mu}$ acts only on the variables of the nucleon $i$.
By Eq.(\ref{eq.1bcurrent}), one can express the hadron tensor as the sum of 
two terms, i.e.,
\begin{equation}
W^{\mu\mu}_{\rm{tot}}(\omega,q) = W^{\mu\mu}(\omega,q) +
W^{\mu\mu}_{\rm{coh}}(\omega,q) \ , \label{eq.sumhadten}
\end{equation}
where $W^{\mu\mu}(\omega,q)$ is the incoherent hadron tensor \cite{west},
which contains only the diagonal contributions
$j_i^{\mu\dagger}Gj_i^{\mu}$, whereas the coherent hadron tensor 
$W^{\mu\mu}_{\rm{coh}}(\omega,q)$ gathers the residual terms 
of interference between different nucleons. As the
incoherent hadron tensor, also $W^{\mu\mu}_{\rm{coh}}(\omega,q)$ can be
expressed in terms of single-particle quantities (see Sect. 9 of Ref.
\cite{capma}), but for the transferred momenta considered here we can take 
advantage of the high-$q$ approximation \cite{orlandini} and retain only 
$W^{\mu\mu}(\omega,q)$.
This term can be further simplified using the symmetry of $G$ for the exchange
of nucleons and the antisymmetrization of $\mid\Psi_0\rangle$. Therefore,
we write
\begin{equation}
W^{\mu\mu}_{\rm{tot}}(\omega,q)  \simeq W^{\mu\mu}(\omega,q) 
=-\frac{A+1}{\pi} {\rm{Im}}  \langle \Psi_0
\mid j^{\mu\dagger}(\q) G(E_{\rm {f}}) j^{\mu}(\q) \mid \Psi_0 \rangle \ ,
\label{eq.apphadrten}
\end{equation}  
where $j^{\mu}(\q)$ is the component of $J^{\mu}(\q)$ related to an arbitrarily
selected nucleon.

Let $\mid n\rangle$ and $\mid \varepsilon \rangle$ denote the 
eigenvectors of the residual Hamiltonian $H_R$ of A interacting nucleons 
related to the discrete and continuous eigenvalues
$\varepsilon_n$ and $\varepsilon$, respectively. We introduce the 
operators $P_n$, projecting
onto the $n$-channel subspace of $\mathcal{H}$, and $Q_n$, projecting onto the
orthogonal complementary subspace, i.e.,
\begin{equation}
P_n = \int \diff \r \mid \r ; n \rangle \langle n;\r \mid ,\, \, \,
Q_n = 1 - P_n  \ . \label{eq.pro}
\end{equation}
Here $\mid \r ;n\rangle$ is the vector obtained from the tensor
product between the discrete eigenstate $\mid n \rangle$ of $H_R$, and the
orthonormalized eigenvectors $\mid \r \rangle$ of the selected nucleon. 
Moreover, we introduce the projection operator onto the continuous channel
subspace, i.e.,
\begin{equation}
P_c = \int \diff \varepsilon \int \diff \r \mid \r ; \varepsilon \rangle 
\langle \varepsilon;\r \mid \ . \label{eq.proc}
\end{equation}
Due to the completeness of the set $\left\{ \mid \r ; n \rangle , 
\mid \r  ; \varepsilon \rangle \right\}$, one has
\begin{equation}
\sum_n P_n + P_c = 1 \ . \label{eq.sumcom}
\end{equation}
Then, we insert Eq. (\ref{eq.sumcom}) into Eq. (\ref{eq.apphadrten}) 
disregarding the contribution of $P_c$. This approximation, which simplifies 
the calculations, is correct for sufficiently high values of the transferred
momentum $q$. Thus, the hadron tensor of Eq. (\ref{eq.apphadrten}) can be
expressed as the sum 
\begin{equation}
W^{\mu\mu}(\omega,q) = W^{\mu\mu}_{\rm{d}}(\omega,q) +
W^{\mu\mu}_{\rm{int}}(\omega,q) \ , \label{eq.sumhad}
\end{equation}
of a direct term
\begin{eqnarray}
W^{\mu\mu}_{\rm{d}}(\omega,q) & = & \sum_n W^{\mu\mu}_{n}(\omega,q) \ ,\nonumber \\
W^{\mu\mu}_{n}(\omega,q)   & = & -\frac{A+1}{\pi} {\rm{Im}} \langle \Psi_0
\mid j^{\mu\dagger}(\q) P_n G(E_{\rm {f}}) P_n 
j^{\mu}(\q) \mid \Psi_0 \rangle \ , \label{eq.directterm}
\end{eqnarray}   
and of a term
\begin{eqnarray}
W^{\mu\mu}_{\rm{int}}(\omega,q)  & = &
\sum_n \widehat W^{\mu\mu}_{n}(\omega,q) \ , \nonumber \\
\widehat W^{\mu\mu}_{n}(\omega,q)  
& = &  -\frac{A+1}{\pi}{\rm{Im}} \langle 
\Psi_0 \mid j^{\mu\dagger}(\q) P_n G(E_{\rm {f}}) Q_n 
j^{\mu}(\q) \mid \Psi_0 \rangle \ ,      \label{eq.intterm}
\end{eqnarray}   
which gathers the contributions due to the interference between the 
intermediate states $\mid \r ;n\rangle$ related to different channels. 

If we disregard the effects of interference between different
channels and consider only the direct contribution to the hadron tensor of 
Eq. (\ref{eq.directterm}), the matrix elements of $P_nG(E)P_n$ in the
basis $\mid \r ; n\rangle$ define the single-particle Green's function 
$\mcg_n(E)$ \cite{capuzzi} of the single-particle Feshbach optical potential 
$\mcv_n(E)$ \cite{fesh} 
\begin{equation}
{\mcg_n(E)} = \frac{1}{E-T-{\mcv_n(E)}+{\rm{i}} \eta} \ , \label{eq.greenn}
\end{equation}
which describes the elastic scattering of a nucleon by
an A-nucleus in the discrete state $\mid n \rangle$. Using the definition of 
$P_n $ in Eq. (\ref{eq.pro}), the direct 
hadron tensor $W^{\mu\mu}_{n}(\omega,q)$ of Eq. (\ref{eq.directterm}) can be 
reduced to the single-particle expression
\begin{equation}
W^{\mu\mu}_{n}(\omega,q) = -\frac{1}{\pi} \lambda_n {\rm{Im}} \langle 
\varphi_n\mid j^{\mu\dagger}(\q) \mcg_n(E_{\rm {f}}-\varepsilon_n) 
j^{\mu}(\q) \mid \varphi_n 
\rangle \ , \label{eq.directht}
\end{equation}
where $\lambda_n$ is the spectral strength \cite{bofficapuzzi} of the hole 
state $\varphi_n $, which is the normalized overlap integral between 
$\mid \Psi_0\rangle$ and $\mid n \rangle$. 
The problem of expressing the interference hadron
tensor $\widehat W_n^{\mu\mu}$ in a one-body form is treated in Ref.
\cite{capuzzi}. It is argued that the contribution of $\widehat W_n^{\mu\mu}$ 
can be included into the direct hadron tensor $W_n^{\mu\mu}$ by the simple 
replacement
\begin{equation}
\mcg_n(E) \rightarrow \mcg_n^{\rm{eff}}(E) \equiv \sqrt{1-\mcv'_n(E)}
\mcg_n(E) \sqrt{1-\mcv'_n(E)} \ , \label{eq.simple}
\end{equation}
where $\mcv'_n(E)$ is the energy derivative of the Feshbach optical potential.

In principle, different optical potentials  $\mcv_n$ must be considered for
different values of $n$. As neither microscopic nor empirical calculations are 
available for the optical potential $\mcv_n$ associated with the excited 
states, a common practice relates them to the ground state potential $\mcv_0$ 
by means of an appropriate energy shift.  Therefore we set
\begin{equation}
\mcv_n(E) \simeq \mcv_0(E) \ , \label{eq.po}
\end{equation}
which implies
\begin{equation}
\mcg_n(E) \simeq \mcg_0(E) \ . \label{eq.pog}
\end{equation}
Using these approximations, we write
\begin{equation}
W^{\mu\mu}(\omega , q) = -\frac{1}{\pi}\sum_n \lambda_n \, {\rm{Im}} 
\langle \varphi_n
\mid j^{\mu\dagger}(\q) \mcg_0^{\rm{eff}} (E_{\rm {f}}-\varepsilon_n)
j^{\mu}(\q) \mid \varphi_n \rangle \ . \label{eq.weff}
\end{equation}

As a next step, the spectral representation of the single-particle
Green's  function function can be used in order to allow practical 
calculations of the hadron tensor of Eq. (\ref{eq.weff}).   

In expanded form, the hadron tensor reads
\begin{eqnarray}
W^{\mu\mu}(\omega , q) = &-& \frac{1}{\pi} \sum_n \lambda_n \, {\rm{Im}} 
\langle \varphi_n
\mid j^{\mu\dagger}(\q) \sqrt{1-\mcv'(E)} \nonumber \\
&\times& \mcg(E)\sqrt{1-\mcv'(E)}j^{\mu}(\q) \mid \varphi_n \rangle \ , 
\label{eq.exweff}
\end{eqnarray}
where $E=E_{\rm {f}}-\varepsilon_n$. Here and below, the lower index $0$ is
omitted in the Green's functions and in the related quantities.

Due to the complex nature of $\mcv(E)$ the spectral representation of $\mcg(E)$
involves a biorthogonal expansion in terms of the eigenfunctions of $\mcv(E)$ 
and of its Hermitian conjugate $\mcv^{\dagger}(E)$. We consider the incoming 
wave scattering solutions of the eigenvalue equations, i.e.,
\begin{eqnarray} 
\left({\mathcal{E}} -  T -\mcv^{\dagger}(E)\right) \mid
\chi_{\mathcal{E}}^{(-)}(E)\rangle &=& 0 \ , \label{eq.inco1} \\ 
\left({\mathcal{E}} - T -\mcv(E)\right) \mid
\tilde {\chi}_{\mathcal{E}}^{(-)}(E)\rangle &=& 0 \ . \label{eq.inco2}
\end{eqnarray}
The choice of incoming wave solutions is not strictly necessary, but it is
convenient in order to have a closer comparison with the treatment of the
exclusive ($e,e'N$) reactions, where the final states fulfill this asymptotic
condition and are the eigenfunctions
$\mid\chi_E^{(-)}(E)\rangle$ of $\mcv^{\dagger}(E)$.

The eigenfunctions of Eqs. (\ref{eq.inco1}) and (\ref{eq.inco2}) satisfy the
biorthogonality condition
\begin{equation}
\langle\chi_{\mathcal{E}}^{(-)}(E)\mid
\tilde {\chi}_{\mathcal{E}'}^{(-)}(E)\rangle = \delta 
\left(\mathcal{E} - \mathcal{E}' \right) \ , \label{bicon}
\end{equation}
and, in absence of bound eigenstates, the completeness relation
\begin{equation}
{\int_M^{\infty}} {\diff} {\mathcal{E}} \mid 
{\tilde{\chi}}_{\mathcal{E}}^{(-)}(E)\rangle\langle
\chi_{\mathcal{E}}^{(-)}(E)\mid 
= 1 \label{eq.comple}
\end{equation}
where the nucleon mass $M$ is the threshold of the continuum of the Feshbach
Hamiltonian. 

Eqs. (\ref{bicon}) and (\ref{eq.comple}) are the mathematical basis for 
the biorthogonal expansions. The contribution of possible bound state solutions 
of Eqs. (\ref{eq.inco1}) and (\ref{eq.inco2}) can be disregarded in 
Eq. (\ref{eq.comple})  
since their effect on the hadron tensor is negligible at the energy and momentum
transfers considered in this paper.

Using Eqs. (\ref{eq.comple}) and (\ref{eq.inco2}), one obtains the spectral 
representation
\begin{equation}
\mcg(E) = {\int_M^{\infty}} {\diff} {\mathcal{E}}\mid 
{\tilde{\chi}}_{\mathcal{E}}^{(-)}(E)\rangle 
 \frac{1}{E-{\mathcal{E}}+{\rm{i}}{\eta}} \langle 
 \chi_{\mathcal{E}}^{(-)}(E)\mid \ . \label{eq.sperep}
\end{equation}
Therefore, Eq. (\ref{eq.exweff}) can be written as
\begin{equation}
W^{\mu\mu}(\omega , q) = -\frac{1}{\pi} {\sum_n}  {\rm{Im}} \bigg[
 {\int_M^{\infty}} {\diff} {\mathcal{E}} \frac{1}{E_{\rm{f}}-
 \varepsilon_{n}-{\mathcal{E}}+{\rm{i}}\eta} T_n^{\mu\mu}({\mathcal{E}} 
 ,E_{\rm{f}}- \varepsilon_n) \bigg]
\ , \label{eq.pracw}
\end{equation}
where
\begin{eqnarray}
T_n^{\mu\mu}({\mathcal{E}} ,E) &=& \lambda_n \langle \varphi_n
\mid j^{\mu\dagger}(\q) \sqrt{1-\mcv'(E)}
\mid\tilde{\chi}_{\mathcal{E}}^{(-)}(E)\rangle \nonumber \\ 
&\times& \langle\chi_{\mathcal{E}}^{(-)}(E)\mid  \sqrt{1-\mcv'(E)} j^{\mu}
(\q)\mid \varphi_n \rangle  \ . \label{eq.tprac}
\end{eqnarray}
The limit for $\eta \rightarrow +0$, understood before the integral of 
Eq. (\ref{eq.pracw}), can be calculated exploiting the standard symbolic 
relation
\begin{equation}
{\lim_{\eta \rightarrow 0}} \, {\frac{1}{E-{\mathcal{E}}+{\rm{i}}\eta}} = 
{\mathcal{P}} \left( {\frac{1}{E-{\mathcal{E}}}} \right) - {\rm{i}} \pi 
\delta \left(E-{\mathcal{E}}\right) \ , \label{eq.princ}
\end{equation}
where ${\mathcal{P}}$ denotes the principal value of the integral. Therefore, 
Eq. (\ref{eq.pracw}) reads
\begin{eqnarray}
W^{\mu\mu}(\omega , q) &=& {\sum_n} \Bigg[ {\rm{Re}} \, T_n^{\mu\mu}
(E_{\rm{f}}-{\varepsilon_n}, E_{\rm{f}}-{\varepsilon_n}) \nonumber \\  
&-& \frac{1}{\pi} {\mathcal{P}}  {\int_M^{\infty}} {\diff} {\mathcal{E}} 
\frac{1}{E_{\rm{f}}-{\varepsilon_n}-{\mathcal{E}}} 
{\rm{Im}} \,T_n^{\mu\mu} ({\mathcal{E}},E_{\rm{f}}-{\varepsilon_n}) \Bigg] \ , 
\label{eq.finale}
\end{eqnarray}
which separately involves the real and imaginary parts of $T_n^{\mu\mu}$.

Some remarks on Eqs. (\ref{eq.tprac}) and (\ref{eq.finale}) are in order.  
Disregarding the square root correction due to interference effects, one
observes that in Eq. (\ref{eq.tprac}) the second matrix element (with the
inclusion of $\sqrt{\lambda_n}$) is the transition amplitude for the single
nucleon knockout from a nucleus in the state $\mid \Psi_0\rangle$ leaving the
residual nucleus in the state $\mid n \rangle$. The attenuation of its strength,
mathematically due to the imaginary part of $\mcv^{\dagger}$, is related to the
flux lost towards the channels different from $n$. In the inclusive response
this attenuation must be compensated by a corresponding gain due to the flux
lost, towards the channel $n$, by the other final states asymptotically
originated by the channels different from $n$. In the description provided by
the spectral representation of Eq. (\ref{eq.finale}), the compensation is
performed by the first matrix element in the right hand side of 
Eq. (\ref{eq.tprac}), where the imaginary part of $\mcv$ has the effect of 
increasing the strength. Similar considerations can be made, on the purely 
mathematical ground, for the integral of Eq. (\ref{eq.finale}), where the 
amplitudes involved in $T_n^{\mu\mu}$ have no evident physical meaning as 
${\mathcal{E}}\neq E_{\rm{f}}-\varepsilon_n$. 
We want to stress here that in the Green's function approach it is just the 
imaginary part of $\mcv$ which accounts for the redistribution of the strength 
among different channels.

The matrix elements in Eq. (\ref{eq.tprac}) contain the mean field $\mcv(E)$ 
and its Hermitian conjugate $\mcv^{\dagger}(E)$, which are nonlocal operator 
with a possibly complicated matrix structure. Neither microscopic nor empirical 
calculations of $\mcv(E)$ are available. In contrast, phenomenological optical 
potentials are available. They are obtained from fits to experimental data, are 
local and involve scalar and vector components only. The necessary replacement 
of the mean field by the empirical optical model potential is, however, a 
delicate step.

In the nonrelativistic treatment of Refs. \cite{capuzzi,capuzzi2} this
replacement is justified on the basis of the approximated equation (holding for
every state $\mid \psi\rangle$)
\begin{eqnarray}
& &  {\rm{Im}} \langle \psi\mid  {\sqrt{1-\mcv'(E)}} {\mcg(E)} 
{\sqrt{1-\mcv'(E)}} \mid \psi\rangle \nonumber \\ 
 &\simeq& {\rm{Im}} \langle \psi\mid {\sqrt{1-\mcv_{\rm {L}}'(E)}} 
 {\mcg_{\rm {L}}(E)} {\sqrt{1-{\mcv_{\rm {L}}'(E)}}} \mid \psi\rangle \ ,\,
\label{eq.approximate}
\end{eqnarray}
where $\mcv_{\rm {L}}(E)$ is the local phase-equivalent potential identified 
with the phenomenological optical model potential and $\mcg_{\rm {L}}(E)$
is the related Green's function. We have reasonable confidence that 
Eq. (\ref{eq.approximate}) holds also in the relativistic context. 

\section{Results}
\label{sec.results}
The response functions of the inclusive QE electron scattering are calculated 
from the single-particle expression of the coherent hadron tensor in  
Eq. (\ref{eq.finale}). After the replacement of the mean field $\mcv(E)$ by the 
empirical optical model potential $\mcv_{\rm {L}}(E)$, the matrix elements of 
the nuclear current operator in Eq. (\ref{eq.tprac}) are of the same kind as 
those giving the transition amplitudes  of the electron induced nucleon 
knockout reaction in the distorted wave impulse approximation (DWIA)
\cite{Oxford}. Thus, calculations have been performed adopting the same 
relativistic \cite{meucci1} and nonrelativistic treatments \cite{DWEEPY} 
which were successfully applied to describe exclusive ($e,e'p$) data, and with
the same phenomenological ingredients for bound state wave functions and optical
potentials.

In the calculations the residual nucleus states $\mid n \rangle$ are restricted 
to be single particle one-hole states in the target. A pure shell model is 
assumed for the nuclear structure, i.e., we take a unitary spectral strength 
for each single particle state and the sum runs over all the occupied states.
In this way we are able to account for the contributions of all the nucleons in
the nucleus, but correlations are completely neglected. This is of course an
approximation that anyhow allows us to perform relatively simple calculations on
a conceptuallly clear basis.

The hadron tensor in Eq. (\ref{eq.finale}) is the sum of two terms. The 
calculation of the second term, which requires integration over all the 
eigenfunctions of the continuum spectrum of the optical potential, is a rather 
complicate task. The contribution of this term is very small and can be 
neglected in the nonrelativistic frame \cite{capuzzi}, while in the relativistic
frame it can be significant and must included in the calculation \cite{rel}.
 
\begin{figure}[ht]
\begin{center}
\includegraphics[height=100mm,width=100mm]{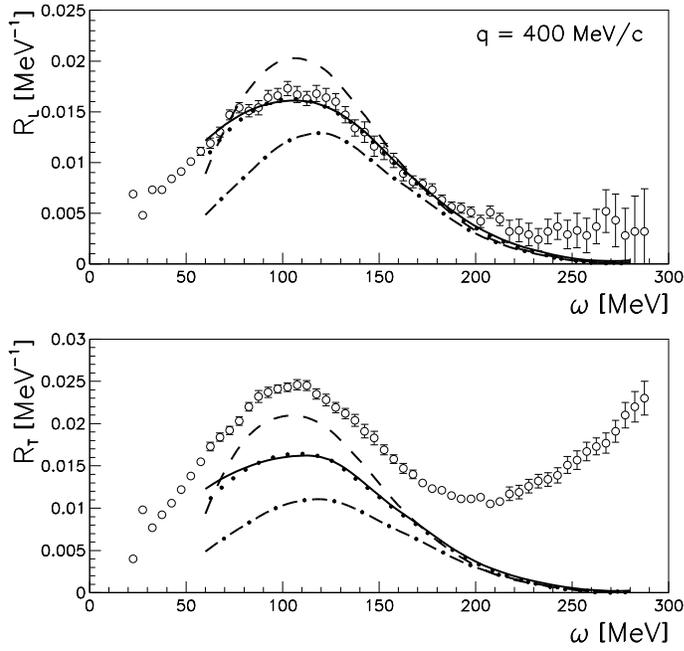}
\vspace{2mm}\caption[]{Longitudinal (upper panel) and transverse (lower panel) 
response functions for the $^{12}$C($e,e'$) reaction at $q = 400$ MeV$/c$ 
calculated in the relativistic frame.  
Solid and dotted lines are obtained with and without the factor 
$\sqrt{1-\mcv'(E)}$, respectively. Dashed lines give the 
result without the 
integral in Eq. (\ref{eq.finale}). Dot-dashed lines are the contribution of 
integrated single nucleon knockout only. The data are from Ref. \cite{saclay}.
(from Ref. \cite{rel})
\label{fig:incl1}
}
\end{center}
\end{figure}

The longitudinal and transverse response functions calculated in the 
relativistic frame for $^{12}$C at $q = 400$ MeV$/c$ are displayed in 
Fig. \ref{fig:incl1} and compared with the Saclay data \cite{saclay}. The low 
energy transfer values are not given because the relativistic optical 
potentials are not available at low energies.

The agreement with the data is generally satisfactory for the $R_{\rm L}$. 
In contrast, $R_{\rm T}$ is underestimated. This
is a systematic result of both relativistic and nonrelativistic calculations. 
It may be attributed to physical effects which are not considered in the 
present approach, e.g., two-body currents. 
The effect of the integral in Eq. (\ref{eq.finale}) is also displayed.
At variance with the nonrelativistic result \cite{capuzzi}, this contribution 
is important and essential to reproduce the experimental longitudinal 
response. The contribution of interference between different channels, that
produces the factor $\sqrt{1- \mcv'_{\rm{L}}(E)}$, gives only a negligible
contribution in Fig. \ref{fig:incl1}, while in the nonrelativistic calculation 
this contribution is important to improve the agreement with $R_{\rm L}$ data 
\cite{capuzzi}.

The contribution from all the integrated single nucleon knockout
channels is also drawn in Fig. \ref{fig:incl1}. It is significantly smaller than 
the complete calculation. The reduction, which is larger at lower values of
$\omega$, gives an indication of the relevance of inelastic channels.  

\begin{figure}[ht]
\begin{center}
\includegraphics[height=100mm,width=100mm]{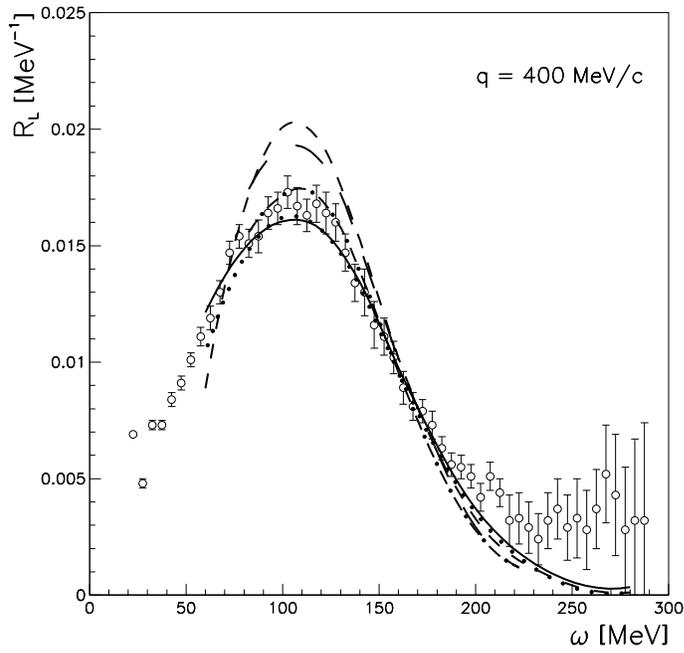}
\vspace{2mm}\caption[]{Longitudinal response function for the
$^{12}$C($e,e'$) reaction at $q = 400$ MeV$/c$.  
Solid dotted and dashed lines are the same as in the upper panel of Fig. 1. 
Dot-dashed and long-dashed lines are the nonrelativistic results with and 
without the factor $\sqrt{1-\mcv'(E)}$, respectively. 
Data as in Fig.1.
(from Ref. \cite{rel})
\label{fig:incl2}
}
\end{center}
\end{figure}

An example of the comparison between the results of the relativistic and the
nonrelativistic approaches is presented in Fig. \ref{fig:incl2} for the
longitudinal response at $q = 400$ MeV$/c$. Both complete calculations are in
satisfactory agreement with the data. This result is however due to a
different effect of the various contributions in the two approaches. In the
nonrelativistic case the factor $\sqrt{1- \mcv'_{\rm{L}}(E)}$ produces a 
substantial reduction of the calculated response, that is necessary to 
reproduce the data. The integral in the second term of  Eq. (\ref{eq.finale})
gives only a small contribution and is neglected in the calculation. In the
relativistic approach the integral is essential to reproduce the data while the
interference between different channels is generally negligible.  

\begin{figure}[ht]
\begin{center}
\includegraphics[height=100mm,width=100mm]{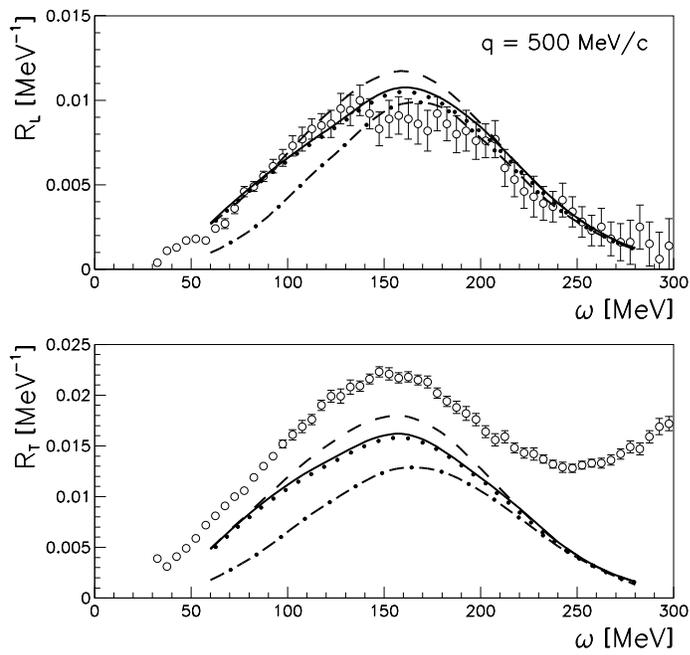}
\vspace{2mm}\caption[]{The same as in Fig. \ref{fig:incl1}, but for $q = 500$ 
MeV$/c$. The data are from Ref. \cite{saclay}. 
(from Ref. \cite{rel})
\label{fig:incl3}
}
\end{center}
\end{figure}

The longitudinal and transverse response functions calculated in the
relativistic approach for $^{12}$C at $q = 500$ are displayed in 
Fig. \ref{fig:incl3}  and compared with the Saclay data \cite{saclay}. 
As already found in Fig. \ref{fig:incl1} at $q = 400$ MeV$/c$, a good agreement 
with the data is obtained for $R_{\rm L}$, while $R_{\rm T}$ is 
underestimated. Also in Fig. \ref{fig:incl3} only a slight effect is 
given by the interference between different channels. The role of the integral 
in Eq. (\ref{eq.finale}) decreases increasing the momentum transfer. 
The effect of the inelastic channels, indicated in the 
figure by the difference between the complete results and the contribution 
from all the integrated single nucleon knockout channels, is always visible 
and even sizable, but it decreases increasing the momentum transfer.

\begin{figure}[ht]
\begin{center}
\includegraphics[height=100mm,width=100mm]{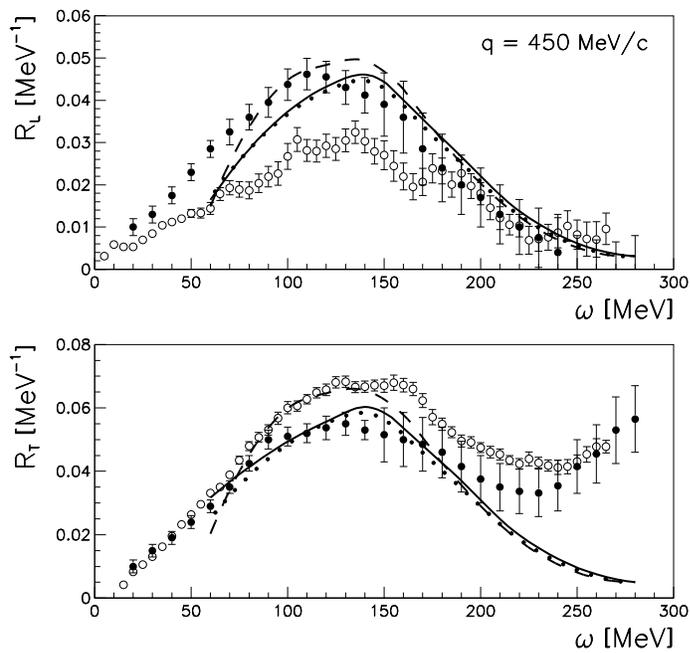}
\vspace{2mm}\caption[]{Longitudinal (upper panel) and transverse (lower panel) 
response functions for the
$^{40}$Ca($e,e'$) reaction at $q = 450$ MeV$/c$. The Saclay
data (open circles) are from Ref. \cite{meziani}, the MIT-Bates (black circles)
are from Ref. \cite{batesca}. Line convention as in Fig. \ref{fig:incl1}.
(from Ref. \cite{rel})
\label{fig:incl4}
}
\end{center}
\end{figure}

The response functions for $^{40}$Ca at $q = 450$ MeV$/c$ 
are shown in Fig. \ref{fig:incl4} and compared with the available data. 
The calculated results are of the same order of magnitude as the 
MIT-Bates data, while for the Saclay data $R_{\rm L}$ is 
overestimated and $R_{\rm T}$ underestimated. 
The factor $\sqrt{1- \mcv'_{\rm{L}}(E)}$ produces and enhancement which is 
minimal but visible in the figure.

\section{Conclusions}
\label{sec.conc}

A Green's function approach to inclusive QE electron scattering has been 
presented. The components of the hadron tensor are written in terms of  
Green's functions of the optical potentials related to the various reaction 
channels. 
The projection operator formalism is used to derive this result. An explicit 
calculation of the single-particle Green's function can be avoided by means of 
its spectral representation, based on a biorthogonal expansion in terms of the 
eigenfunctions of the non-Hermitian optical potential $\mcv(E)$ and of its 
Hermitian conjugate. The interference between different channels is taken into 
account by the factor $\sqrt{1-\mcv'(E)}$, which also allows the replacement of 
the mean field $\mcv(E)$ by the phenomenological optical potential 
$\mcv_{\textrm{L}}(E)$. After this replacement, the nuclear response functions 
are expressed in terms of matrix elements similar to the ones which appear in 
the exclusive one nucleon knockout reactions, and the same DWIA treatment can 
be applied to the calculation of the inclusive electron scattering. 

The effects of FSI are thus described consistently in exclusive and inclusive 
processes. Both the real and imaginary parts of the optical potential must be 
included. In the exclusive reaction the imaginary part accounts for the flux 
lost towards other final states. In the inclusive reaction, where all the final 
states are included, the imaginary part accounts for the redistribution of the 
strength among the different channels. 

All the final states contributing to the inclusive reaction are contained in 
the Green's function, and not only one nucleon emission. Our calculations for 
the inclusive electron scattering are different from the contribution of 
integrated single nucleon knockout only. The difference between the two results 
is originated by the imaginary part of the optical potential.

FSI effects are similar on the longitudinal and transverse components of the 
nuclear response and are important both in the relativistic and in the 
nonrelativistic calculations. The role and relevance of the various effects can 
however be different in the two frames. The final effect is similar and 
produces qualitatively similar results in comparison with data. The 
longitudinal response is usually well reproduced, while the transverse response 
is underestimated. This seems to indicate that more complicated effects, e.g., 
two-body meson exchange currents, have to be added to the present single 
particle approach.

%%%%%%%%%%%%%%
%%
%% FIGURES
%%
%%%%%%%%%%%%%%
%%%%%%%%%%%%%%

\end{document}